 \documentclass[aps,prl,twocolumn,superscriptaddress,showpacs,floatfix]{revtex4}
\usepackage{graphicx}
\usepackage{dcolumn}
\usepackage{bm}
\begin{document}

\title{Hydrogen-induced disintegration of fullerenes and nanotubes:\\
       An \textit{ab initio} study }

\author{Savas Berber}
\affiliation{Physics and Astronomy Department,
             Michigan State University,
             East Lansing, Michigan 48824-2320, USA}
\affiliation{Physics Department,
             Gebze Institute of Technology,
             Gebze, Kocaeli 41400, Turkey}

\author{David Tom\'anek}
\affiliation{Physics and Astronomy Department,
             Michigan State University,
             East Lansing, Michigan 48824-2320, USA}

\date{\today}


\begin{abstract}
We use {\em ab initio} density functional calculations to study
hydrogen-induced disintegration of single- and multi-wall carbon
fullerenes and nanotubes. Our results indicate that hydrogen atoms
preferentially chemisorb along lines in sp$^2$ bonded carbon
nanostructures, locally weakening the carbon bonds and releasing
stress. For particular structural arrangements, hydrogen helps to
relieve the accumulated stress by inducing step-wise local
cleavage leading to disintegration of the outermost wall.
\end{abstract}




\maketitle




Exposure to hydrogen is known to cause embrittlement in
metals\cite{DT067}, a major concern in materials science. It is an
intriguing question, whether local decohesion due to chemisorbed
hydrogen may cause similar damage in sp$^2$ bonded carbon
nanostructures, such as fullerenes and
nanotubes\cite{Dresselhaus96}, and whether it may have helped to
efficiently cleave graphene\cite{Dairibbon08}. So far, past
research has focussed on beneficial side-effects of the
interaction of hydrogen with nanocarbons, including easier
separation/debundling and the prospect of reversible hydrogen
storage\cite{{Dresselhaus96},{Jorio08}}. Since both fullerenes and
nanotubes are under internal stress due to the local curvature,
hydrogen-induced local decohesion may facilitate fracture and
cause irreversible damage in nanocarbons as it does in
metals\cite{DT067}.


Here we present an {\em ab initio} density functional study of
bonding and structural changes in sp$^2$ carbon nanostructures
exposed to hydrogen. Our results indicate that hydrogen atoms
preferentially chemisorb along lines on the outermost wall of
single- and multi-wall carbon fullerenes and nanotubes, locally
weakening the carbon bonds and releasing stress. For particular
arrangements of adsorbed hydrogen atoms, we identify the
microscopic process of step-wise local cleavage leading to the
disintegration of the outermost wall as a way to release the
accumulated stress.


To gain fundamental insight into the effect of hydrogen on the
structural integrity of sp$^2$ bonded carbon nanostructures, we
studied the total energy and optimum structure of pristine and
hydrogenated fullerenes and nanotubes. Our results are based on
the density functional theory (DFT) within the local density
approximation (LDA) and first-principles pseudopotentials. We used
primarily the SIESTA\cite{soler2002} code with a double-$\zeta$
basis set with polarization orbitals, the
Perdew-Zunger\cite{Perdew81} parametrization of the
exchange-correlation functional, and norm-conserving {\em ab
initio} pseudopotentials\cite{Troullier91} in their fully
separable form\cite{Kleinman82}. The accuracy of our results was
verified using the PWSCF plane-wave code with ultrasoft
pseudopotentials. We used $10^{-2}$~eV/{\AA} as a strict gradient
convergence criterion when determining optimum geometries.
Constrained structure optimization has been used to investigate
energy barriers, transition states, and generally to identify the
optimum reaction pathway.


Published theoretical results include studies of the bonding
character of isolated H atoms on graphite flakes, nanotubes and
fullerenes\cite{jeloaica99,DT193,DT198} and the preferential
arrangement of hydrogen pairs on a graphene
layer\cite{{allouche06},{roman07},{hornekaerPRL06}}. Other
investigations address the relative stability of hydrogen pairs
for different adsorption geometries \cite{Crespi03} and
coverages\cite{allouche06b} of hydrogen on graphene. Theoretical
studies of hydrogen-covered nanotubes focussed on
adsorbate-related electronic structure changes\cite{GulserenPRB02}
and the relative stability of different adsorption
arrangements\cite{{Yildirim01},{arellano02},{Crespi03},{DT193},{GulserenPRB03}}.
More recent studies considered the possibility of axial cleavage
in nanotubes in the presence of exohedrally adsorbed H
atoms\cite{{LuPRB03},{arellano02},{DT193}}. No results have been
reported so far about the origin of hydrogen-induced structural
damage in fullerene-like structures and its relation to
stress-induced cleavage of nanotubes.


Same as in graphene, each carbon atom in fullerenes and nanotubes
has three neighbors to form very stable sp$^2$ bonds. Unlike in
planar graphene, the most stable sp$^2$ carbon allotrope, stress
is associated with the nonzero local curvature in fullerenes and
nanotubes. The small fullerenes C$_{20}$ and C$_{60}$ are
spherical, with all atoms equivalent and subject to the same
stress. With increasing size, however, free-standing fullerenes
start resembling icosahedra, with twenty near-planar triangular
facets spanning the surface. In these structures, stress
associated with higher local curvature concentrates near the
thirty edges connecting adjacent pentagons. In carbon nanotubes,
on the other hand, all atoms are equivalent and the stress
associated with local curvature is distributed uniformly
throughout the structure.

\begin{figure}[t]
\includegraphics[width=0.9\columnwidth]{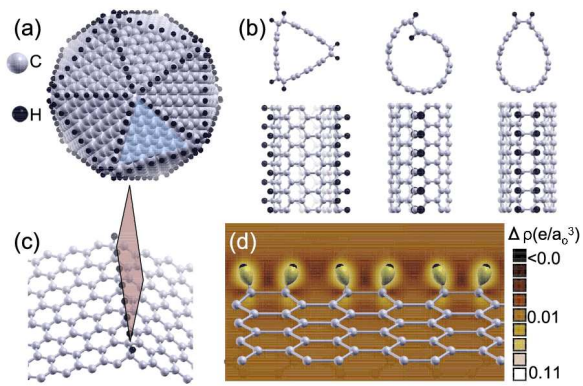}
\caption{(Color online) Equilibrium structure and bonding
character in H-covered fullerenes and nanotubes. (a) Hydrogen
chemisorption on the C$_{960}$ fullerene along strained lines with
larger curvature. (b) Lines of chemisorbed hydrogen, shown in
end-on and side view, yield the most stable adsorption on the
(6,6) carbon nanotube. (c) Relaxed atomic structure near a line of
chemisorbed hydrogen atoms. (d) Difference charge density
$\Delta{\rho}$ indicating hydrogen-induced
sp$^2\!\!{\rightarrow}$sp$^3$ local bonding change in graphitic
nanocarbons. Contours of $\Delta{\rho}$ in the plane indicated in
(c) are superimposed with the atomic lattice.\label{Fig1}}
\end{figure}


Even though fullerenes with more than 100 atoms have not been
observed, large icosahedral shells are believed to form the
outermost shell of multi-wall fullerenes, called bucky
onions\cite{Ugarte-onion-Nature92}. Focussing on the faceted
C$_{960}$ hollow shell structure in Fig.~\ref{Fig1}(a), increased
stress makes edge atoms more reactive and prone to bind hydrogen.
Molecular hydrogen does not dissociate on planar graphene, but
does so with an energy gain on fullerenes, with the dissociated
hydrogen pair preferentially binding on top of adjacent carbon
atoms. In the following, we will define adsorption energies per
hydrogen pair, related to the free H$_2$ molecule. As expected,
the energy gain depends on the local stress in the substrate, with
values ranging from $1.20$~eV at the twelve pentagonal corners to
$1.35$~eV at the edge sites and $0.34-0.67$~eV on the facets of
the C$_{960}$ fullerene.


Calculations at higher coverages suggest that after the first two
atoms are adsorbed, additional hydrogens should preferentially
adsorb along straight lines on the substrate\cite{DT193}. We
believe that formation of this energetically preferred adsorption
pattern can be achieved by surface diffusion of hydrogen atoms at
temperatures found during many hydrogenation reactions, since the
activation barrier for sigmatropic rearrangement of chemisorbed
hydrogen atoms is only ${\approx}1$~eV\cite{DT193}. As suggested
earlier and illustrated in Fig.~\ref{Fig1}(c), carbon atoms at
icosahedral edges in large fullerene cages are more reactive than
the facet sites. Since the dissociative chemisorption energy of
hydrogen along the edges exceeds that on fullerene facets by
${\approx}0.7$~eV per H$_2$ molecule, we expect hydrogen atoms to
adsorb preferentially along lines connecting pentagon corners.


Even though there are no preferential adsorption sites on a carbon
nanotube, atomic hydrogen is believed to adsorb preferentially
along lines parallel to the tube
axis\cite{{LuPRB03},{arellano02},{DT193}} to release stress.
Possible initial adsorption geometries for hydrogen on the $(6,6)$
carbon nanotube are shown in Fig.~\ref{Fig1}(b). In the left
panel, we depict the optimum nanotube geometry following hydrogen
adsorption along three zigzag lines parallel to the axis. The
calculated binding energy of 0.54~eV per hydrogen pair reflects
also the energy associated with the tube relaxation to a
triangular cross-section. Presence of hydrogen apparently locally
reduces the flexural rigidity of the nanotube wall, converting its
circular cross-section into a triangle.

Other favorable local adsorption patterns include exo- or
endohedrally adsorbed hydrogens forming double-lines along the
nanotube axis. The initial geometry of such a combined
exo/endohedral adsorption arrangement is shown in the middle
panel, and an exohedral double-line arrangement of hydrogens is
shown in the right panel of Fig.~\ref{Fig1}(b).
As we discuss later on, hydrogen-induced local decohesion causes
significant structural rearrangements in both cases. The energy
gain associated with hydrogen chemisorption was found to be
$0.87$~eV per hydrogen pair in this case.


To obtain fundamental insight into the origin of hydrogen-induced
local decohesion that is general and independent of a particular
graphitic structure, we studied the effect of a line of
chemisorbed hydrogen atoms on the structure and bonding in a
graphene layer, as shown in Fig.~\ref{Fig1}(c). Hydrogen atoms
bind preferentially on top of carbon atoms, causing a local
buckling. A line of chemisorbed hydrogen atoms is expected to form
a crease in the initially planar graphene\cite{Dairibbon08}. In
Fig.~\ref{Fig1}(c) we constrained the crease angle to equal that
between adjacent facets on the icosahedron in Fig.~\ref{Fig1}(a).
To get insight into the nature of the adsorption bond, we
inspected the charge density difference ${\Delta}{\rho}=\rho{\rm
(H/graphene)}-\rho{\rm (H)}-\rho{\rm (graphene)}$, representing
the charge redistribution in the system upon hydrogen adsorption.
The spatial distribution of ${\Delta}{\rho}({\bf r})$ is shown in
Fig.~\ref{Fig1}(d) in the plane schematically indicated in
Fig.~\ref{Fig1}(c). Hydrogen adsorption causes only local charge
rearrangement leading to a charge accumulation in the H-C $\sigma$
bonds, depleting the charge in the local C-C bond region, as
expected in case of sp$^3$ bonding. Both the charge rearrangement
and the local buckling are strong indicators for a local
sp$^2{\rightarrow}$sp$^3$ transition. As suggested above, hydrogen
pairs bind preferentially to neighboring carbon atoms, thereby
changing an initially strong sp$^2$ bond to a weaker, strained
sp$^3$ bond. This local decohesion resembles to some degree the
microscopic mechanism of hydrogen embrittlement in some
metals\cite{DT067}.


Besides decohesion, local change from sp$^2$ to sp$^3$ bonding is
also associated with a net increase in the C-C bond length by
typically 8\%. In partially hydrogen-covered systems, carbon atoms
connected to lines of hydrogen are under stress. The consequences
can be best illustrated in the fullerene structure shown in
Fig.~\ref{Fig1}(a), where hydrogen only attaches to the edges. The
hydrogen-free planar facets can be thought as incompressible, same
as small graphene flakes. Since the corners of the triangular
facets determine the edge length, sp$^3$ bonded carbon atoms along
the edges are prevented from increasing their interatomic
distance, subjecting the edges to stress. The energy associated
with accumulated stress increases linearly with the edge length
$L$. For a critical value $L_{crit}$, this energy will exceed the
activation barrier ${\Delta}E$ to break a C-C bond as a first step
to initiating a cleavage that eventually causes the disintegration
of the structure.

As we show later on, we find that ${\Delta}E{\approx}1.7$~eV. This
amount of stress energy is accumulated in a hydrogenated fullerene
edge of critical length $L_{crit}{\approx}15$~nm. We thus expect
hydrogen-induced cleavage to occur only in fullerenes with
$L>15$~nm, with a diameter exceeding ${\approx}28$~nm and
containing more than $10^5$ C atoms. Such large fullerenes never
occur as single-wall structures, but may form the outermost wall
of a multi-wall bucky onion. In large bucky onions, the spherical
walls inside exert radial pressure on the outermost wall, thus
enhancing the hydrogen-induced stress and reducing the critical
diameter for disintegration of the outermost wall. We expect that
multi-wall fullerenes exposed to hydrogen should peel
layer-by-layer from the outside in a process appearing as
exfoliation.

\begin{figure}[b]
\includegraphics[width=0.75\columnwidth]{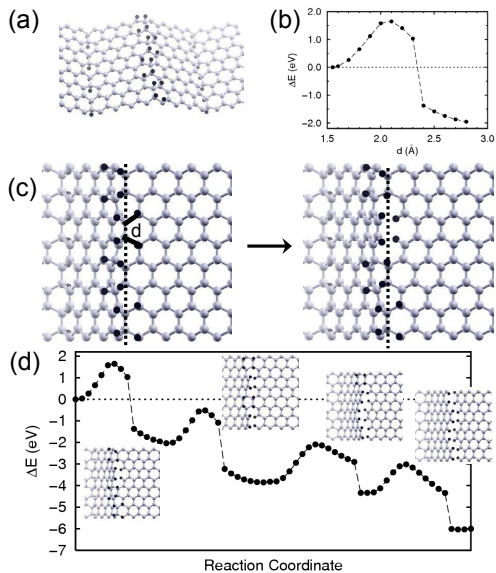}
\caption{Zipper mechanism of hydrogen-induced disintegration of
strained sp$^2$ carbon nanostructures. (a) Periodically warped
graphene as model strained system. (b) Optimum structure and
energetics for breaking a bond pair. (c) Sequence of bond
cleavages resulting in a crack. \label{Fig2}}
\end{figure}


In the following, we study the microscopic mechanism leading to
cleavage at the minimum energy cost. To obtain a general
understanding of the fracture process that does not depend on a
particular system, we created generic creases by placing a
periodic array of hydrogen lines along the armchair direction on
alternating sides of an infinite graphene monolayer, as seen in
Fig.~\ref{Fig2}(a). The energetics and atomic arrangement during
the initial cleavage are depicted in Figs.~\ref{Fig2}(b) and
\ref{Fig2}(c). Cleavage at the crease is caused by locally
weakening the affected bonds. We found that this is best achieved
by adsorbing additional hydrogen atom pairs on both sides of the
initial hydrogen line, as shown in the left panel of
Fig.~\ref{Fig2}(c). Such a local arrangement of adsorbed hydrogens
is a favorable prerequisite to cleave an edge segment. We found
that the energy needed to break one bond is larger than the energy
to break two adjacent bonds, shown in a dark color in the left
panel of Fig.~\ref{Fig2}(c). Continuing this concerted bond
breaking mechanism should cleave the edge, causing the structure
to disintegrate.


The energetics of cleavage in Fig.~\ref{Fig2}(b) shows the total
energy change with respect to the initial structure as a function
of the bond length $d$ across the crack, which we use as the
reaction coordinate. We consider the full range of $d$ values
occurring during the cleavage reaction and, for each value of $d$,
we globally optimize the structure. Consequently, the calculated
transition paths for the disintegration of hydrogenated fullerenes
and nanotubes represent the optimum cleavage path in unconstrained
systems. Following an initial energy investment of $1.7$~eV, the
bonds break abruptly, leading to the final structure depicted in
the right panel of Fig.~\ref{Fig2}(c). There is a net energy gain
of ${\approx}2$~eV associated with the bond breaking, caused by
releasing the accumulated stress. The initially stressed sp$^3$
bonded crease segment converts into a pair of more stable sp$^2$
bonded, hydrogen terminated graphene edges.


We next calculated the energy barriers associated with breaking
specific bonds in search of the optimum pathway to complete the
crease cleavage after it was initiated. The optimum sequence of
bonds to break, along with the intermediate structures and the
energetics of the process, is depicted in Fig.~\ref{Fig2}(d).
Interestingly, the smallest energy investment to propagate the
fracture following the first step in Fig.~\ref{Fig2}(b) does not
involve the neighboring pair of C-C bonds on the opposite side of
the ridge, but rather the more distant pair of C-C bonds on the
same side of the ridge. Our results indicate that the energy
barrier to propagate the fracture decreases at successive steps.
As a matter of fact, it may be reasonable to assume that the
energy released following the initial step may activate the
zipper-like cleavage at the crease without further energy
investment. Once cleavage is initiated, the unzipping process
transforms the stressed crease into two overlapping graphene edges
in an exothermic reaction.

\begin{figure}[t]
\includegraphics[width=0.9\columnwidth]{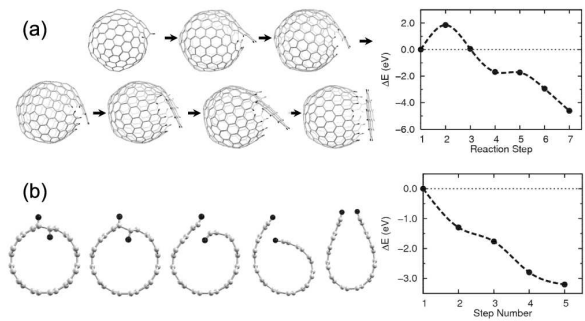}
\caption{Hydrogen-assisted disintegration of (a) the C$_{240}$
fullerene and (b) the (6,6) carbon nanotube. Structural snap shots
along the optimum reaction pathway are depicted in the left
panels. The structural transformations of the nanotube in (b) are
shown in end-on view. Carbon atoms are shown as grey, hydrogen
atoms as larger black spheres. The right panels show the total
energy change in the system, with the data points corresponding to
the structures in the left panels. \label{Fig3}}
\end{figure}


Having studied the microscopic cleavage process in the warped
graphene model, we next turn to hydrogen-induced disintegration of
specific fullerene and nanotube structures. As we discussed above,
we expect hydrogen-induced disintegration to proceed spontaneously
only in very large fullerenes, which are typically multi-wall
structures. In the smaller C$_{240}$ fullerene, we find this
reaction to be energetically activated. The energetics of the
C$_{240}$ disintegration process, induced by hydrogen adsorption
along the icosahedral edges, is shown in Fig.~\ref{Fig3}(a) along
with snap-shots of intermediate structures. Unlike in the model
system addressed in Fig.~\ref{Fig2}, the hydrogen coverage has
been gradually increased as the cleavage progressed, as an
alternative scenario to initiating unzipping. Both in the warped
graphene model and in the C$_{240}$ fullerene, we find the
disintegration process to be exothermic, following an initial
energy investment of ${\approx}2$~eV.


Whereas large fullerenes are more susceptible to hydrogen-induced
fracture due to the stress build-up in the hydrogenated edges, we
find nanotubes that are narrow to be more susceptible to fracture
due to the stress caused by their high uniform curvature. Both in
fullerenes and nanotubes with multiple walls, hydrogen will first
adsorb on the outermost wall, causing local decohesion leading to
bond breaking and eventual disintegration of the outermost wall.
Continuing exposure to hydrogen will cause a layer-by-layer
peeling that will appear as exfoliation.


As suggested by our results in Fig.~\ref{Fig1}(b), hydrogen
adsorption concentrates the deformation along lines parallel to
the tube axis, reducing the flexural stress in the hydrogen-free
region. The right panels of Fig.~\ref{Fig1}(b), depicting hydrogen
covered nanotubes, illustrate how hydrogen-induced decohesion
along adsorption lines may cause large structural changes. The
detailed cleavage process of an infinitely long $(6,6)$ carbon
nanotube, covered by a complete chain of hydrogens adsorbed
exohedrally and a neighboring chain adsorbed endohedrally, is
described in Fig.~\ref{Fig3}(b). The sequence of snap shots,
presented in an end-on view, depicts intermediate structures,
including a scroll\cite{DT138}, encountered during a conjugate
gradient optimization from an unrelaxed initial structure. This
process is only spontaneous if all hydrogens adsorb in the optimum
way to form completely filled lines of adsorbed hydrogen, and is
activated otherwise.
The final structure with a horseshoe-like cross-section,
stabilized by the attraction of the hydrogenated zigzag edges, may
furthermore undergo an activated conversion to a planar graphene
strip with hydrogen-terminated edges.


In conclusion, we used {\em ab initio} density functional
calculations to study hydrogen-induced disintegration of single-
and multi-wall carbon fullerenes and nanotubes. Our results
indicate that hydrogen atoms preferentially chemisorb along lines
in sp$^2$ bonded carbon nanostructures, locally weakening the
carbon bonds and releasing stress. Contrary to common belief that
chemisorbed hydrogen causes no harm to carbon nanostructures, we
show that irreversible structural changes may be caused by
chemisorbed hydrogen that can initiate local cleavage leading to
the disintegration of the outermost wall. In experimental
observations of multi-wall structures, this could be interpreted
as hydrogen-induced exfoliation.


We thank Glen P. Miller for inspiring this study by his
experimental observations in bucky onions. This work was funded by
the National Science Foundation under NSF-NSEC Grant No.~425826
and NSF-NIRT Grant No.~ECS-0506309. Computational resources have
been provided by the Michigan State University High Performance
Computing Center.


\end{document}